\documentclass{article}
\usepackage[preprint]{spconfa4}
\usepackage{spconfa4,amsmath,amssymb,graphicx,url,times}
\usepackage{color}
\usepackage{flushend}
\usepackage{url}
\usepackage[sort]{cite}
\usepackage{subcaption}
\copyrightnotice{978-1-5386-8151-0/18/\$31.00~\copyright2018 IEEE}

\title{Simulating Multi-Channel Wind Noise Based on the Corcos Model}

\twoauthors{Daniele Mirabilii}{}{Emanu\"{e}l A.P. Habets}{} 
\address{International Audio Laboratories Erlangen*,  Am Wolfsmantel 33, 91058 Erlangen, Germany\\
	\{daniele.mirabilii,emanuel.habets\}@audiolabs-erlangen.de \thanks{*A joint institution of the Friedrich-Alexander-University Erlangen-N\"urnberg (FAU) and Fraunhofer IIS, Germany.}}

\begin{document}

\ninept
\maketitle

\begin{sloppy}

\begin{abstract}
A novel multi-channel artificial wind noise generator based on a fluid dynamics model, namely the Corcos model, is proposed. In particular, the model is used to approximate the complex coherence function of wind noise signals measured with closely-spaced microphones in the free-field and for time-invariant wind stream direction and speed. Preliminary experiments focus on a spatial analysis of recorded wind noise signals and the validation of the Corcos model for diverse measurement set-ups. Subsequently, the Corcos model is used to synthetically generate wind noise signals exhibiting the desired complex coherence. The multi-channel generator is designed extending an existing single-channel generator to create $N$ mutually uncorrelated signals, while the predefined complex coherence function is obtained exploiting an algorithm developed to generate multi-channel non-stationary noise signals under a complex coherence constraint. Temporal, spectral and spatial characteristics of synthetic signals match with those observed in measured wind noise. The artificial generation overcomes the time-consuming challenge of collecting pure wind noise samples for noise reduction evaluations and provides flexibility in the number of generated signals used in the simulations.
\end{abstract}


\section{Introduction}
\label{sec:intro}
When a free-field air stream encounters an obstacle (such as a microphone), the solid geometry causes a transition from a laminar to a turbulent flow. Turbulences in the layer of the air in the immediate proximity of a bounding surface, known as the \textit{turbulent boundary layer} (TBL), result in wind noise in microphone recordings. This wind noise often degrades outdoor recordings of speech, whose quality and intelligibility are impaired by low-frequency rumbling artifacts. For this reason, different wind noise reduction algorithms were developed in the past decades, using single-channel \cite{nelke2014single,nemer2013single,hofmann2012morphological} or multi-channel \cite{Thune2016,elko2007reducing,franz2010multi,nelke2014dual} approaches. Existing multi-channel algorithms are most commonly derived under the assumption that the wind noise is spatially uncorrelated, implying a zero-valued complex coherence function. 
   
For the development and performance evaluation of wind noise reduction algorithms, a controlled environment is required: a large variety of wind noise samples must be added to clean speech samples with a predefined input SNR, to measure the achieved noise suppression. The challenging task of wind noise capturing motivates the design of synthetically generated signals: outdoor measurements are often disrupted by diverse acoustic sources (e.g., traffic, natural ambient sounds etc.) from which pure wind noise is difficult to isolate. Furthermore, indoor recordings require low reverberant rooms and the air stream source used in the experiments (e.g., the fan motor) is also included in the recordings. For this purpose, a wind noise database and a single-channel artificial wind noise generator was presented in \cite{nelke2014measurement}, based on measured wind noise temporal and spectral statistics.

In this contribution, we analyze the complex coherence of wind noise signals at different inter-microphone distances. Our analysis shows a non-zero complex coherence for smaller inter-microphone distances. Based on experimental results, we show that the complex coherence of wind noise signals can be approximated by a theoretical model when the inter-microphone distance is small and the wind direction and velocity are constant. A variety of models have been identified to approximate the stochastic spatial pressure distribution associated to convective turbulences in the TBL. Among others, the Corcos model \cite{corcos1964structure} is well-established in the aerodynamics literature for low Mach fluid stream speed.

The Corcos model is used in this work to approximate the complex coherence function of wind noise contributions in multi-channel simulations and to generate noise signals exhibiting the desired complex coherence. First, the single-channel generator from \cite{nelke2014measurement} is extended to create $N$ spatially white wind noise signals. Secondly, the algorithm proposed in \cite{habets2008generating} is exploited to compute an instantaneous mixing matrix through a decomposition of the complex coherence matrix defined by the Corcos model. Finally, the $N$ wind noise signals are mixed using the aforementioned matrix to obtain the desired complex coherence.  

The remainder of this paper is organized as follows. In Section~\ref{sec:MCA} we present a spatial analysis on recorded wind noise signals, we introduce the Corcos model and we use the latter to approximate the complex coherence of the measured wind noise signals. In Section~\ref{sec:WNS} we use the Corcos model to artificially generate wind noise signals with the desired complex coherence. In Section \ref{sec:validation} the complex coherence of generated wind noise signals is compared to the theoretical model using the normalized mean squared error. Finally, in Section \ref{sec:conclusions} we draw some conclusion based on the obtained results.

\section{Spatial Coherence Analysis}
\label{sec:MCA}
\subsection{Coherence of Measured Wind Noise}

\begin{figure}[!t]
	\centering
	\begin{subfigure}{1\linewidth} 
		\includegraphics[width=\linewidth]{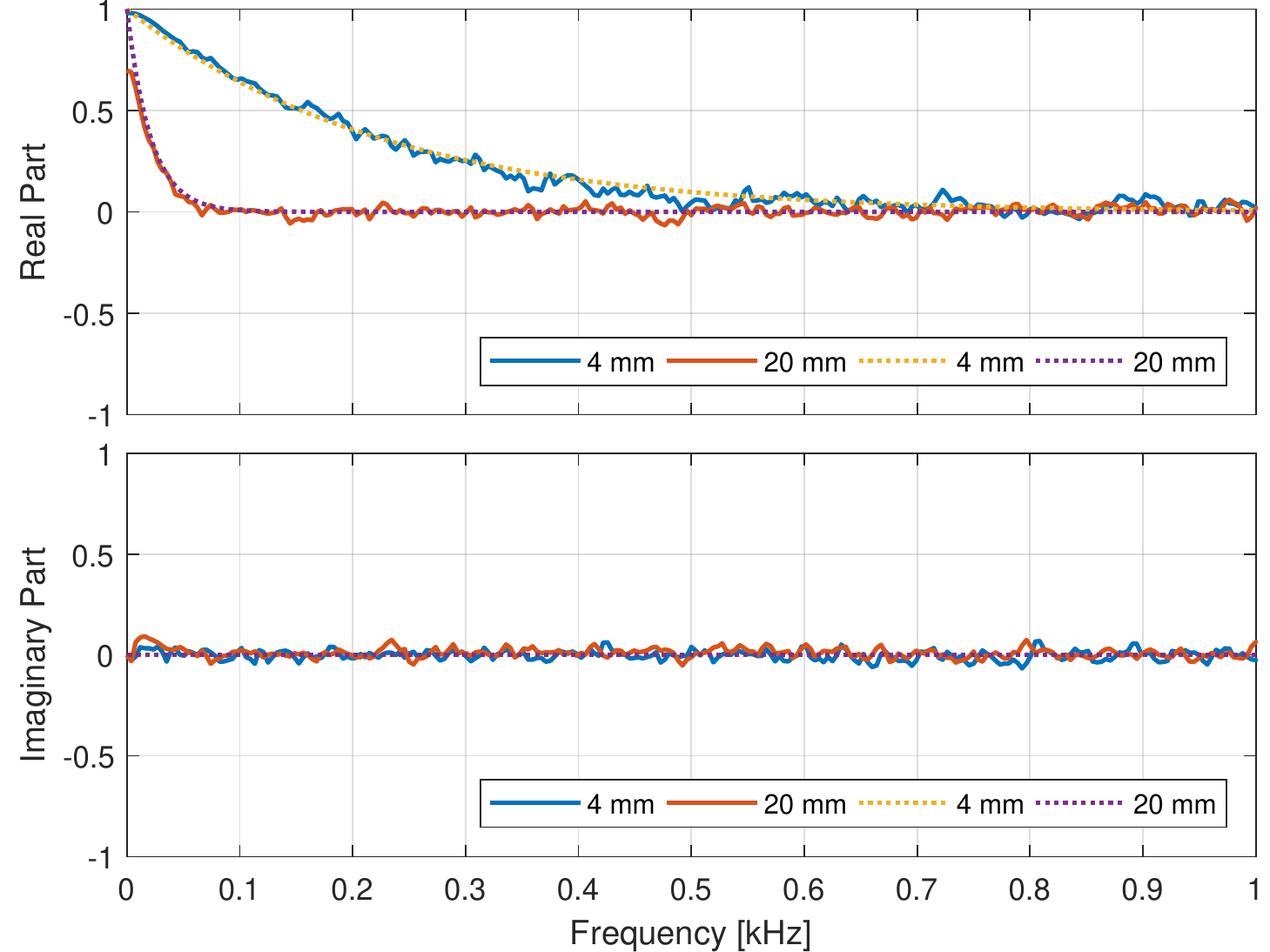}
		\caption{Crosswind: $\theta_w \approx \pi/2$ rad, $U=1.8$ m/s.  } 
	\end{subfigure}
	\vspace{1em} 
	\begin{subfigure}{1\linewidth} 
		\includegraphics[width=\linewidth]{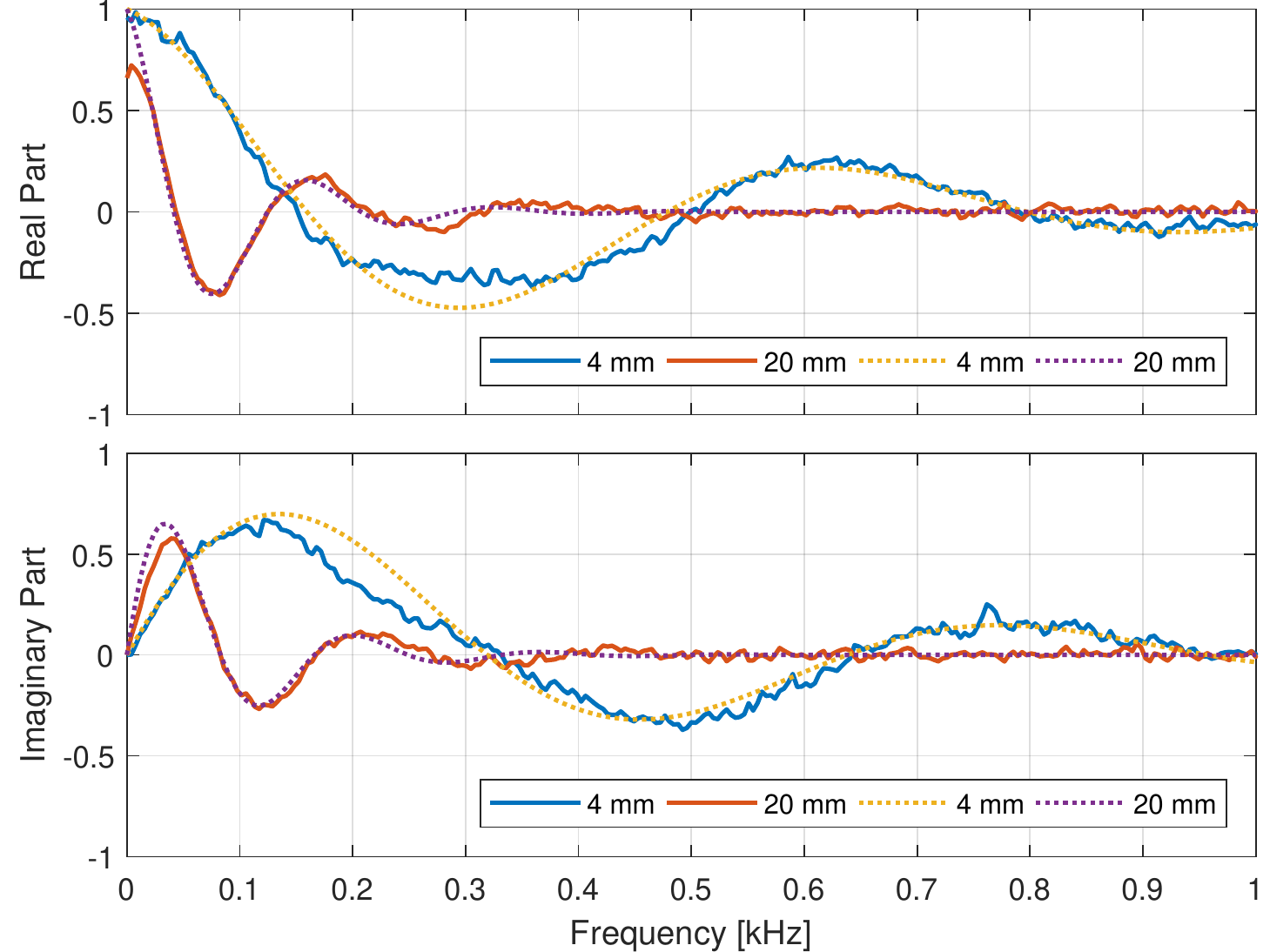}
		\caption{Downwind: $\theta_w \approx 0$ rad, $U=2.8$ m/s.} 
	\end{subfigure}
	\caption{Real and imaginary part of the complex coherence function of different wind noise measurements (solid lines) compared to the Corcos model (dashed lines). } 
	\label{fig:1}
\end{figure}

In this section we analyze the spatial characteristics of wind noise recorded indoors. The reliability of an indoor air stream compared to an outdoor wind flow was provided in \cite{nelke2014measurement}. A further comparison between indoor and outdoor measurements showed similar characteristics in terms of the temporal behaviour, spectral components and power spectral density (PSD) differences between the channels.

The measurement set-up consists of two omnidirectional MEMS microphones exposed to a stationary fan-generated air stream. The microphone are positioned in the free-field of a low-reverberant room. The investigation is carried out for different inter-microphones distance, two fixed directions of the air stream (assuming a constant laminar-flow propagation in the free-field) and two stream velocities which were measured by an anemometer. The two directions of the air stream are denoted by \textit{crosswind} for a flow orthogonal to the microphone axis (in the following $\theta_w=\frac{\pi}{2}$ radians) and \textit{downwind} for a flow parallel to the microphone axis (in the following $\theta_w=0$ radians). The free-field air stream velocity is denoted by $U$ and expressed in m/s.

The solid lines in Fig. \ref{fig:1} depict the real and the imaginary part of the complex coherence function of measured wind noise for different configurations. The complex coherence was computed for a total duration of 120 s of the microphone signals. The analysed frequency range is limited to 1 kHz to avoid the high coherence values of the fan engine between 1.5-7 kHz, which acts like a single acoustic source. Moreover, in \cite{NelkePhD2016}, an investigation on the cumulative energy distribution of wind noise shows that the 99.5 \% of the spectral energy is below 1 kHz. Two main considerations can be formulated from the obtained results: (1) the complex coherence is different from zero under the assumptions of  free-field measurements, constant air stream direction $\theta_w$ and speed $U$ and (2) under these assumptions, the complex coherence of wind noise signals depends on the distance between the microphones, the direction and the speed of the free-field air stream. Increasing the microphone distance results in a faster decay of the complex coherence towards zero, clearly visible in Fig. \ref{fig:1} (a) where the solid blue and red lines depict the complex coherence of measured wind noise with a microphone distance of 4 mm and 20 mm respectively. For a crosswind stream the complex coherence is a pure real function (a), while a downwind stream implies oscillations in both real and imaginary parts (b).

\subsection{Corcos Model}
 The Corcos model \cite{corcos1964structure} approximates the stochastic pressure distribution associated to convective turbulences in a turbulent boundary layer, which cause the excitation of the microphone membrane, i.e., the generation of wind noise. The cross-PSD between two points at distance $d$ in which the pressure field of the turbulent boundary layer is observed (i.e., two microphones) is a function of the similarity variables $\omega d \lvert\cos(\theta_w)\rvert / U_c$ and $\omega d \lvert\sin(\theta_w) \rvert/ U_c$, where $\omega$ denotes the angular frequency, $\theta_w$ denotes the direction of arrival (DOA) of the flow assuming a plane wave propagation and $U_c$ denotes the convective turbulent wind speed in m/s. The latter was empirically determined in \cite{corcos1964structure} as an approximated value of the 80\% of the free-field wind velocity $U$. 
 
 The model is here re-defined for the expression of the complex coherence function of wind noise signals measured with two microphones at distance $d$ in the discrete frequency-domain formulation and using the microphone axis as reference system, as
 \begin{equation}
 \gamma_c(\omega) = \exp{\left ( \frac{-\alpha(\theta_w) \omega_k d}{ U_c }  \right )} \exp{\left ( \frac{\iota \: \omega_k d\cos(\theta_w)}{ U_c }  \right )}, \label{eq:1} 	
 \end{equation} 
 with $\omega_k = 2\pi k F_s/K$ denoting the discrete angular frequency, where $k$ denotes the frequency bin index, $K$ denotes the length of the discrete Fourier transform and $F_s$ denotes the sampling frequency, while $\alpha(\theta_w)$ denotes a coherence decay parameter that depends on the stream direction, defined by 
\begin{equation}
\alpha(\theta_w)=\alpha_1\lvert\cos(\theta_w)\rvert+\alpha_2\lvert\sin(\theta_w)\rvert,\label{eq:2}
\end{equation} where $\alpha_1$ and $\alpha_2$ are the longitudinal and the lateral coherence decay rates respectively,  experimentally computed in \cite{mellen1990modeling}. Therefore, the pressure field of an air stream loses coherence with frequency-dependent longitudinal (parallel to the stream direction) and lateral (orthogonal to the stream direction) exponential decays, showing higher magnitude coherence values for very low frequency ranges. From (\ref{eq:1}) it is clear that, for larger distances between the microphones, the coherence decay results in a faster decrease toward zero. The phase term in (\ref{eq:1}) causes low frequency oscillations in both real and imaginary part for wind stream DOAs that maximize the quantity $\cos(\theta_w)$, i.e., for downwind streams with $\theta_w = 0$ radians, while the complex coherence becomes a pure real function for crosswind streams with $\theta_w = \pi/2$ radians.

In Fig. \ref{fig:1} the theoretical model (dashed lines) is compared to the measured data (solid lines) for both crosswind (a) and downwind (b) streams. Under the aforementioned assumptions, the measured data follows the theoretical curves and thus the Corcos model, which provides an approximation of the complex coherence of wind noise signals, can be employed for artificial generation.

\section{Multi-Channel Wind Noise Simulation}
\label{sec:WNS}
Simulated wind noise signals picked up by an uniform linear array of microphones (ULA) are generated at the output of the proposed algorithm, whose complex coherence is given by the Corcos model. As a starting point, we use the single-channel generator in \cite{nelke2014measurement} to obtain temporal and spectral characteristics similar to measured wind noise. In particular, the source-filter model in \cite{nelke2014measurement} is extended to generate $N$ uncorrelated wind noise signals under given assumptions. The subsequent complex coherence constraint is achieved using the algorithm presented in \cite{habets2008generating}, in which a generation of multichannel non-stationary noise signals that exhibit a predefined complex coherence is performed.

Starting from $N$ statistically independent wind noise signals generated by $N$ instances of the single-channel generator extended with specific temporal and spectral properties between the individual contributions, the signals are subsequently mixed using a frequency-dependent instantaneous mixing matrix to induce the desired complex coherence. The free parameters are the number of microphones signals $N$, the sample frequency $F_s$, the inter-distance of the microphones $d$, the free-field flow velocity $U$, and the air stream direction $\theta_w$.      

\subsection{Modeling Temporal Characteristics}
\label{sec:3.1}
As described in \cite{nelke2014measurement}, a source-filter system provides an excitation signal modulated by a long-term gain and a short-term gain to simulate the temporal characteristics of measured wind noise. The excitation signal is a weighted sum of a Gaussian white noise and a randomly chosen entry selected from an excitation codebook, collected from real wind noise recordings. The long-term and short-term gains respectively model the amplitude envelope and the fine structure of the generated signals.

The long-term gain is computed exploiting a trained three-state Markov model that simulates the occurrence rate of the temporal transitions of wind noise energy (categorized in three classes: no wind, low wind and high wind). Each state of the Markov model identifies a specific value for the long-term gain.

The short-term gain is modelled as the square root of the frame energy, approximated by a realization of a stochastic process exhibiting a Weibull cumulative distribution. Both long-term and short-term gains are subsequently smoothed by Hann windows of different lengths respectively, in order to avoid unnatural sharp fluctuations in the overall gain, especially during state transitions.

For the multi-channel generation, the single-channel generation is performed $N$ times with three assumptions: (1) the excitation signal for each microphone is a different realization of a stochastic process (Gaussian white noise) weighted with a random entry of the excitation codebook. This guarantees spatially white wind noise contributions, (2) the long-term gain must be shared between the $N$ contributions. This is motivated by the same wind stream exposure and therefore by the identical current state of the Markov model for every microphone, leading to the same amplitude envelope of the contributions, (3) the short-term gain applied to each excitation signal is a different realization of a random process: the instantaneous signal levels must differ from one microphone to the other to be consistent with measured wind noise signals.

\subsection{Modeling Spectral Characteristics}
\label{sec:3.2}
The spectral shape of the wind noise can be described using an auto-regressive model. In \cite{nelke2014measurement}, a time-invariant set of AR parameters is extracted from real wind recordings. The minimum order that maximize the prediction gain was found to be 5. The AR coefficients define the filter $A(z)$, whose magnitude response present a low-pass characteristic, reflecting the spectral shape of wind noise. For the multi-channel generation, every modulated excitation signal is filtered by the same all-pole filter $A(z)$, as every contribution must exhibit the same spectral properties.   

\subsection{Modeling Spatial Characteristics}
\label{sec:3.3}
To generate $N$ wind noise contributions with a complex coherence given by the Corcos model, we use the approach presented in \cite{habets2008generating}, where multiple microphone signals with a predefined complex coherence are obtained by multiplying $N$ uncorrelated signals with an instantaneous mixing matrix, retrieved by a Cholesky decomposition of the complex coherence matrix.

Let us define the complex coherence matrix $\mathbf{\Gamma}(l,k) \in \mathbb{C}^{N \times N } $, where $N$ is the number of microphones and $l$ denotes the time frame index of the short-time Fourier transform (STFT). Each entry of the complex coherence matrix $\gamma_{ij}(k)$ represents the complex coherence values between the $i$-th and the $j$-th microphone, denoted by 
\begin{equation}
	\mathbf{\Gamma}(l,k) = 
	\begin{bmatrix}
		\gamma_{11}(k) & \gamma_{12}(k) & \cdots & \gamma_{1N}(k)\\
		\gamma_{21}(k) & \gamma_{22}(k) & \cdots & \gamma_{2N}(k)\\
		\vdots & \vdots & \ddots & \vdots\\
		\gamma_{N1}(k) & \gamma_{N2}(k) & \cdots & \gamma_{NN}(k)\\
	\end{bmatrix} \forall l. \label{eq:3}
\end{equation}
Simulating closely-spaced microphones in an ULA exposed to a wind stream in the free field, each entry of the complex coherence matrix can be approximated by the Corcos model and denoted as 

\begin{equation}
\gamma_{ij}(k) =
\begin{cases}
 \exp{\left ( \frac{\omega_k d  |i-j| (-\alpha(\theta_w)   + \iota  \cos(\theta_w)) }{U_c}  \right )} \quad \textrm{for} \: i \geq j \\ \\
 \exp{\left ( \frac{\omega_k d  |i-j| (-\alpha(\theta_w)   - \iota  \cos(\theta_w)) }{U_c}  \right )} \quad \textrm{for} \: i < j \\
 \end{cases}
 \label{eq:4} 
\end{equation}


Assuming constant convective turbulent speed $U_c$ and air stream DOA $\theta_w$, the coherence matrix is time-invariant. The instantaneous mixing matrix $\mathbf{C}(k)$ is obtained through a Cholesky decomposition of the complex coherence matrix in (\ref{eq:3}). Assuming a Hermitian positive-definite complex coherence matrix, $\mathbf{\Gamma}(k)$ is decomposed into a product of two matrices \begin{equation}
	\mathbf{\Gamma}(k) = \mathbf{C}^\textrm{H}(k) \mathbf{C}(k), \label{eq:eq4.3}
\end{equation} 
where $(\cdot)^\textrm{H}$ denotes the Hermitian operator and $\mathbf{C}(k)$ is an upper triangle matrix. For a positive-definite matrix, the Cholesky decomposition is unique, so that $\mathbf{C}(k)$ has an unique form for any given $\mathbf{\Gamma}(k)$. Given $N$ uncorrelated wind noise contributions $\mathbf{v}(n) = [v_1(n),...,v_N(n)]^T$ generated exploiting the approach described in Section \ref{sec:3.1} and Section \ref{sec:3.2}, and defining a vector containing their respective STFT coefficients $\mathbf{V}(l,k) = [V_1(l,k),...,V_N(l,k)]^T$, the resulting wind noise signals $\mathbf{\widetilde{V}}(l,k) = [\widetilde{V}_{1}(l,k),..., \widetilde{V}_{N}(l,k)]^T$ exhibiting the complex coherence approximated by the Corcos model are obtained by
\begin{equation}
\mathbf{\widetilde{V}}(l,k)= \mathbf{C}^\textrm{H}(k)\mathbf{V}(l,k).  \label{eq:eq4.8}
\end{equation}
The wind noise time signals $\mathbf{\widetilde{v}}(n) = [\widetilde{v}_{1}(n),..., \widetilde{v}_{N}(n)]^T$ are then computed by the inverse STFT of $\mathbf{\widetilde{V}}(l,k)$.

\section{Performance evaluation}
\label{sec:validation}
In this section the spatial characteristics of generated wind noise signals are analysed and compared to the theoretical model. The similarity between measured and synthetic wind noise in terms of temporal and spectral characteristics presented in \cite{nelke2014measurement} is still valid for the proposed multichannel generator.
\begin{figure}[!t]
	\centering
	\begin{subfigure}{1\linewidth} 
		\includegraphics[width=\linewidth]{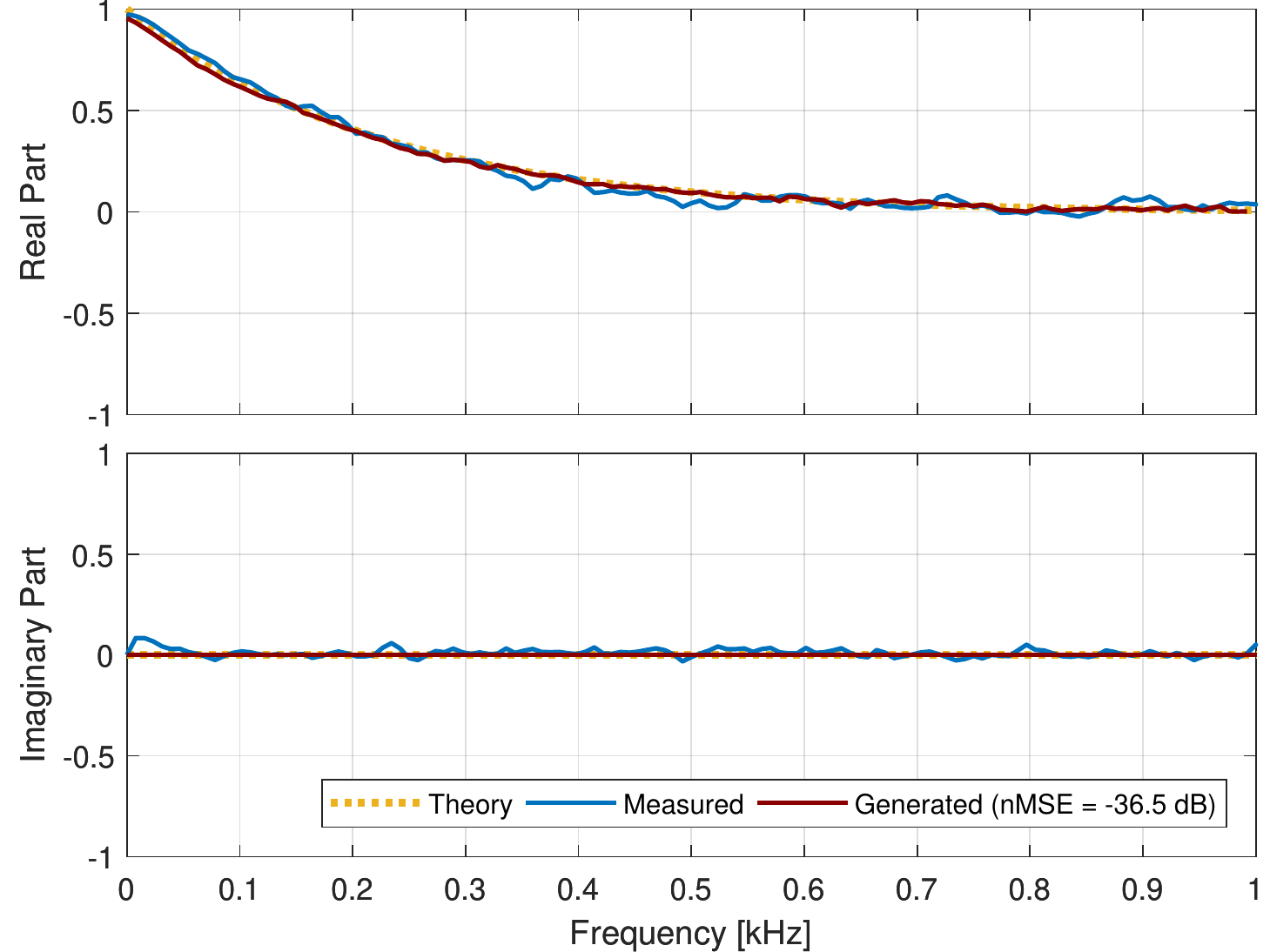}
		\caption{Crosswind: $\theta_w \approx \pi/2$ rad, $U=1.8$ m/s, d = $4$ mm.  } 
	\end{subfigure}
	\vspace{1em} 
	\begin{subfigure}{1\linewidth} 
		\includegraphics[width=\linewidth]{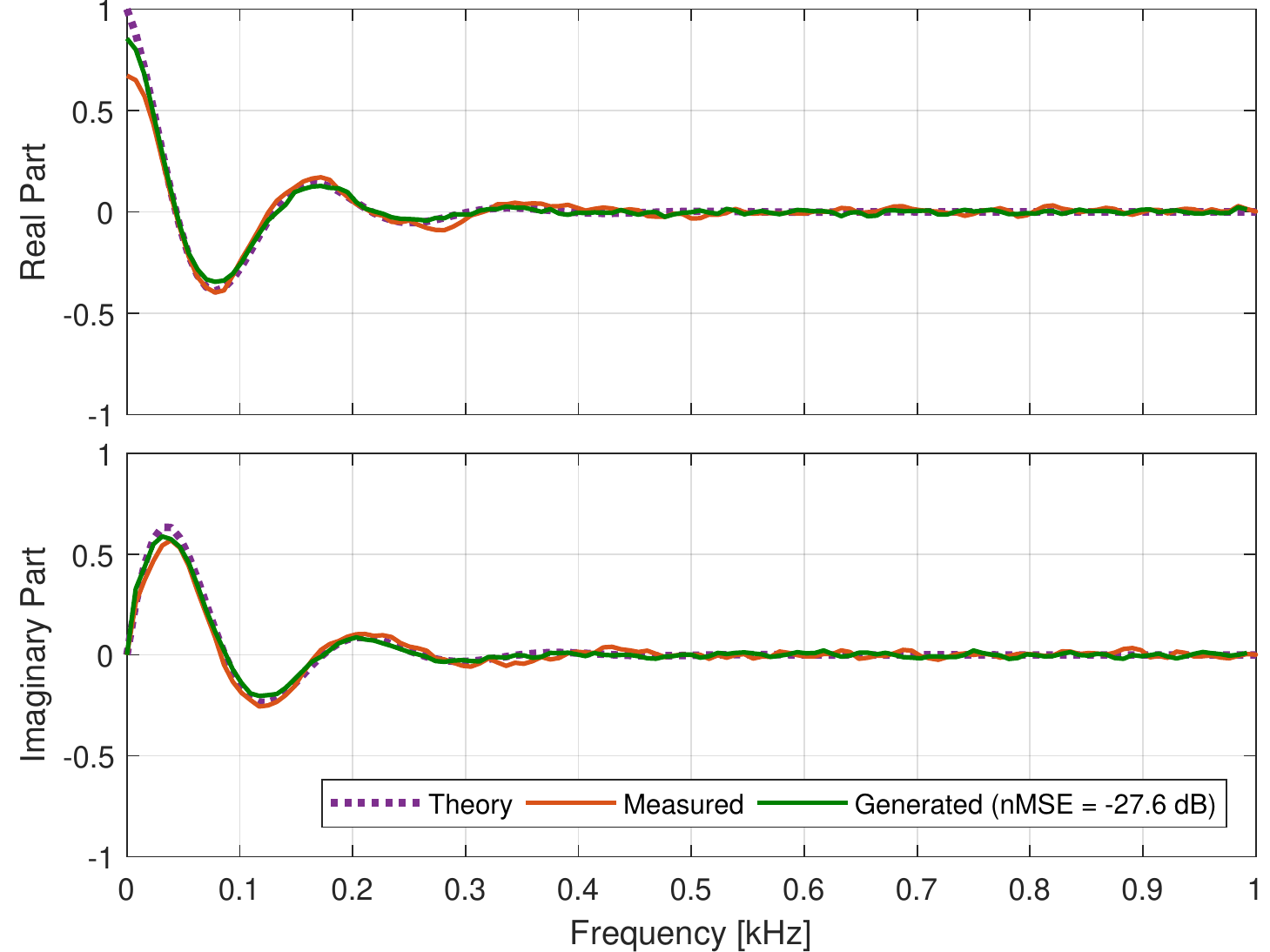}
		\caption{Downwind: $\theta_w \approx 0$ rad, $U=2.8$ m/s, d = $20$ mm.} 
	\end{subfigure}
	\caption{Complex coherence function of generated wind noise signals in comparison to the theoretical model.} 
	\label{fig:2}
\end{figure}

To validate the proposed approach in terms of spatial characteristics, the normalized mean squared error between the theoretical model $\gamma_{ij}(k)$ given by (\ref{eq:4}), and the artificially generated complex coherence $\widehat{\gamma}_{ij}(k)$ obtained by the proposed method is computed for a microphone pair $i$-$j$ as \begin{equation}\textrm{nMSE}_{ij}=\frac{\displaystyle\sum_{k=0}^{K/2-1}|\widehat{\gamma}_{ij}(k)-\gamma_{ij}(k)|^2}{\displaystyle\sum_{k=0}^{K/2-1}|\gamma_{ij}(k)|^2}.
\end{equation}
The generated complex coherence was computed for a total duration of 600 seconds of synthetic wind noise, using a sampling frequency $F_s = 16 $ kHz, an STFT with discrete Fourier transform length of $K=2048$, and a Hann window of $K$ samples (128 ms) with 75\% overlap. We used the same frequency resolution for the definition of the complex coherence matrix in (\ref{eq:3}) and the computation of the STFT coefficients $\mathbf{V}(l,k) = [V_1(l,k),...,V_N(l,k)]^T$ of the uncorrelated wind noise signals.
 
Figure \ref{fig:2} shows the obtained complex coherence in comparison to the theoretical model (dashed lines) given by (\ref{eq:4}) and to the measured data, for two different simulations: (a) a crosswind stream with a microphones distance of 4 mm and (b) a downwind stream with a microphones distance of 20 mm. We chose the mentioned simulations to allow a further comparison with the complex coherence of measured wind noise, shown in Fig. \ref{fig:1} (solid blue line in (a) and solid red line in (b)). The simulated complex coherence follows the theoretical model showing small values of the nMSE (two examples in Fig. \ref{fig:2}) for each simulation. Therefore the generated complex coherence exhibits a close match with both the measured complex coherence and the complex coherence given by the Corcos model.       
 
Wind noise signals generated by the proposed approach and measured signals share comparable temporal, spectral and spatial characteristics. Moreover, informal listening tests resulted in a perceptual satisfaction compared to measured wind noise. The MATLAB implementation of the multi-channel artificial wind noise generator can be found at 
\textit{\url{https://www.audiolabs-erlangen.de/fau/professor/habets/software/noise-generators}}. 

\section{Conclusions}
\label{sec:conclusions}
We developed an algorithm to generate multi-channel wind noise signals exhibiting a complex coherence given by a fluid-dynamics model by Corcos. The latter was defined in the discrete-frequency domain, using the microphone axis as reference system. According to the Corcos model, the complex coherence of wind noise signals depends on the inter-microphone distance, the direction of the air stream and the free-field flow velocity. A preliminary spatial analysis of measured wind noise showed that the Corcos model is a valid approximation of the complex coherence of wind noise contributions measured with closely-spaced microphones exposed to a constant air stream in the free-field. We developed the multi-channel generator as an extension of an existing single-channel generator to obtain $N$ uncorrelated wind noise signals under defined assumptions on the shared temporal and spectral characteristics. The desired complex coherence was subsequently induced by an instantaneous mixing matrix, whose coefficients were obtained through a Cholesky decomposition of the complex coherence matrix, defined by the Corcos model. Multiple parameters can be arbitrarily chosen to generate wind noise signals: the number of microphones $N$, the duration of the signals, the sampling frequency $F_s$, the inter-microphone distance $d$, the air stream direction $\theta_w$ and the free-field flow velocity $U$. The proposed generator can therefore be exploited to create an extensive database of wind noise samples, simulating different scenarios. The simulation of wind noise contributions picked up by an ULA can be further extended to any planar array geometry.  The performance evaluation showed that the complex coherence of synthetically generated wind noise signals closely match the theoretical model and consequently the measured wind noise complex coherence. This demonstrates the feasibility in the usage of artificially generated wind noise for multichannel wind noise reduction simulations, for both objective and subjective evaluation.       

\pagebreak
\flushend
\bibliographystyle{IEEEtran}
\bibliography{IWAENC_2018_SMCWN}

\end{sloppy}
\end{document}